**Structural topography-mediated high temperature wetting symmetry breaking**

Jing Li[1], Youmin Hou[2], Yahua Liu[1], Chonglei Hao[1], Minfei Li[1], Manoj Chaudhury[3, *], Shuhuai Yao[2,*], Zuankai Wang[1,*]

[1]Department of Mechanical and Biomedical Engineering,
City University of Hong Kong, Hong Kong 999077, China

[2]Department of Mechanical and Aerospace Engineering,
The Hong Kong University of Science and Technology, Hong Kong 999077, China

[3]Department of Chemical Engineering,
Lehigh University, Bethlehem, PA 18015

Corresponding authors: zuanwang@cityu.edu.hk (Z.W.) or meshyao@ust.hk (S.Y.) or mkc4@lehigh.edu (M.C.)

**Abstract**

Directed motion of liquid droplets is of considerable importance in various industrial processes. Despite extensive advances in this field of research, our understanding and the ability to control droplet dynamics at high temperature remain limited, in part due to the emergence of complex wetting states intertwined by the phase change process at the triple-phase interfaces. Here we show that two concurrent wetting states (Leidenfrost and contact boiling) can be manifested in a single droplet above its boiling point rectified by the presence of asymmetric textures. The breaking of the wetting symmetry at high temperature subsequently leads to the preferential motion towards the region with higher heat transfer coefficient. We demonstrate experimentally and analytically that the droplet vectoring is intricately dependent on the interplay between the structural topography and its imposed thermal state. Our fundamental understanding and the ability to control the droplet dynamics at high temperature represent an important advance and enable the rational design of various surfaces for multifunctional applications, especially in high temperature thermal systems where high energy efficiency, security and stability are preferred.

Droplets moving on solid surfaces have found important applications in self-cleaning coatings[1], thermal management systems[2, 3], agriculture[4, 5], and microfluidic assay[6]. Droplets deposited or impinged on a uniform hydrophobic surface in the ambient condition usually exhibit an axisymmetric spreading and their motion is also limited due to the contact line pinning[7-11]. By harnessing imposed gradients of surface energy[12-16], light[17], temperature[18], electric force[19, 20], or mechanical vibration[21], the wetting symmetry can be broken and a directional motion is induced. However, despite extensive advances in this field of research, our understanding and the ability to control droplet dynamics at high temperature remain limited, in part due to the emergence of complex wetting states intertwined by the phase change process at the triple-phase interfaces[22-24]. When the temperature of surface is above a critical temperature, a continuous vapor layer separates the droplet from the hot surface[25-30]. This so-called Leidenfrost condition is characterized by a minimal friction as well as a low heat transfer between the droplet and hot solid surface. Crucial to many industrial processes such as spray cooling and fuel injection in combustion engines[31], an effective contact between the impinging droplet and hot solid surface is highly desired. Previous studies have shown that the Leidenfrost point of droplet can be enhanced or suppressed by carefully controlling roughness and wettability of structured surfaces[27, 32-34]. In particular, a liquid droplet can be self-propelled along ratchet structures[35-37], whereas in this case the droplet still maintains in a uniform wetting state and its relevance for practical applications at high temperature systems remains limited.

Here we show that the wetting symmetry normally imposed to an impacting droplet at the room temperature can be broken or retained by judicious control of structural topography and the range of operating temperatures of solid substrate. Particularly, we demonstrate that two concurrent wetting states (Leidenfrost and contact boiling) can be manifested in a single droplet above its boiling point rectified by the presence of asymmetric textures, subsequently leading to the preferential droplet motion from unwanted Leidenfrost regime to boiling regime. Such observation, in some way, is reminiscent of the Buller's droplet effect in nature[38], which is widely exploited by basidiomycete fungi to actively eject their spores. We also develop hydrodynamic models to elucidate the dependence of the Leidenfrost point on structural topography. The utility of asymmetric textures demonstrated here might offer new avenue for enhancing energy transfer and stabilizing energy and high temperature thermal systems including high-density electronics, nuclear power plant under a wider range of temperatures.

The surface with structural roughness gradient (non-uniform texture), consisting of micropost arrays with uniform diameter ($D$ = 20 μm) and height ($H$ = 20 μm), but varying post-to-post spacing ($L$), was fabricated through photolithography and deep reactive-ion etching (see methods). Figure 1a shows the scanning electron microscopy (SEM) image of such an asymmetric surface. The static contact angles (CAs) in different regions are controlled by the post-to-post spacing ($L$) changing from 25 µm in the densest region to 105 µm in the sparsest region over a span of 8.5 mm, resulting in a gradient static CA from 24.8° to 35.7°. Figure 1b and supplementary Fig.1 shows the equilibrium wetting state of a water droplet released at a small Weber number ($We$) of 1.0 on the center of the fabricated surface ($L$ = 65 µm). Here $We = \rho v_o^2 R_o / \gamma$, with $\rho$ and $\gamma$ representing the density and surface tension of the water droplet, and $v_o$ is the impact velocity. The water droplet was ~ 1.42 mm in radius ($R_o$). The droplet reached a stable state at 648 ms, with a slight shift in the centroid by 0.27 mm, indicating the left-right wetting symmetry of the droplet was not very sensitive to structural roughness gradient at the room temperature. The wetting symmetry was maintained at increasing $We$ without pronounced directional motion of droplets (Supplementary Fig. 2).

Substantially different from the imposed wetting symmetry at the room temperature, we observed a preferential droplet motion toward sparser textures at the high temperature (Methods). Figure 2a presents selected snapshots of droplet impacting on the gradient surface at $We$ = 19.3 under $T$ = 245 °C and 265 °C, respectively (Supplementary Movie 1). The droplet reached its maximum spreading diameter ($D_{max}$ ~ 6.8 mm) at 4.4 ms and the leftmost front of the liquid droplet was in contact with sparse posts with $L$ = 95 µm whereas the rightmost front contacted dense posts with $L$ = 35 µm. On retraction, the rightmost front was more mobile, moving radially inwards with a dynamic receding CA greater than 150°. Indeed, close-up view of triple-phase interfaces on dense post arrays (Fig. 2b) reveals the existence of vapor pockets underneath the droplet, which is a typical signature of the Leidenfrost state. By contrast, explosive satellite droplets were observed on the surface with sparse posts, suggesting the manifestation of the contact boiling (CB) state. Thus, the droplet in this situation displayed two distinct wetting states with an asymmetric distribution of vapor layer underneath. We designate this particular state as the mixed boiling-Leidenfrost state (MBL). The directional transport of liquid from the unwanted Leidenfrost regime to the CB regime is relevant in many high temperature systems where the occurrence of a Leidenfrost condition may cause severe problems. Notably, the liquid vectoring achieved in the MBL regime is similar to the ejection of spore by basidiomycetes in nature[38] (Fig. 2c): a water droplet (or Buller's droplet) first nucleates at the base of the spore. When growing large enough, the drop then coalesces with the liquid film around the spore, giving rise to an adequate momentum to propel the spore forward. Comparably, the liquid in the Leidenfrost regime on our surface behaves like the Buller's droplet while the liquid in the CB regime serves as the sticky spore. Distinct from the ejection of spore which is mainly controlled by the surface tension, the preferential droplet motion in our case is mediated by the thermal state of the substrate. This is also different from the droplet motion on very short nanostructured surfaces[39, 40] which exclusively relies on the occurrence of a uniform boiling state.

However, we found that when $T$ was above 295°C or less than 225 °C, the wetting symmetry was maintained and hence there was no preferential droplet motion observed. As shown in Fig. 2d and Supplementary Movie 2, when $T$ was above 295 °C, the droplet stayed in the Leidenfrost state and a symmetric, complete bouncing occurred. The contact time $T_c$ for the droplet to complete one bouncing cycle was ~16 ms, following the scaling of $T_c \sim 2.6\left(\rho D_o^3 / 8\gamma\right)^{1/2}$. This contact time is also independent of the impinging velocity, consistent with that on typical superhydrophobic surfaces at the room temperature[41-43] (Supplementary Fig. 3). When the temperature was less than 225 °C, the droplet stayed in the explosive CB condition (Fig. 2d and Supplementary Movie 2) and vanished within a short time. Taken together, the manifestation of a preferential motion on the gradient surface is dependent on the interplay between the structural topography and its imposed thermal state.

Fig. 3a plots the variation of the maximum displacement of the centroid of the impinging droplet in one impact cycle ($\Delta L$) relative to its initial radius, or $k = \Delta L / R_o$, as a function of substrate temperature. The maximum vectoring occurred at a critical temperature of ~265 °C, and then decayed when the imposed temperature deviated from it. A defining feature of the directional motion is the time for the droplet to completely depart from the Leidenfrost area (Fig. 3b). In the lower temperature range of the MBL regime, the droplet rapidly vectored into the boiling region within 15 ms without droplet bouncing observed. Notably, at the higher temperature range, the droplet manifested an oblique bouncing and it cannot be totally shifted to the CB region within one impact circle. In this condition, the total vectoring time was dramatically extended to vary between 48 ms and 70 ms until the whole droplet cannot be transferred into CB regions when the temperature was too high. By contrast, in both CB and Leidenfrost regimes, the directional motion was inhibited with $k$ close to zero.

We propose that the preferential and rapid droplet motion ensues from the breaking of the symmetry of the triple-phase contact line throughout the droplet. To elucidate how the structural topography is translated into an asymmetric triple-phase contact line, we developed a new dynamic Leidenfrost model. Note that the coupling of dynamic impact process with the complex phase process on the structured surface complicates its analysis far more than what is traditionally applicable for a stationary Leidenfrost droplet[44]. As the droplet impacts on the hot surface, it is subject to a dynamic pressure ($0.5\rho v_o^2$) and a hammer pressure ($\Delta P_h$)[45, 46] which tend to drive the droplet wicking the substrate. $\Delta P_h$ is expressed as $\sim k_h \rho v_o v_s$, where $v_s$ is the speed of sound in water and the value of the pre-factor $k_h$ can be calculated by the surface morphology[47]. On the other hand, upon contacting superheated posts, the liquid evaporates in the vicinity of posts while an outward vapor pressure gradient is developed[48]. Based on the previous model[49], the thickness of the vapor layer above posts $\delta$ was estimated to be ~ 1 μm in our experiment. Thus, the vapor flow can be divided into two layers: (1) the thin vapor layer above post arrays; (2) the vapor flow inside post arrays (Supplementary Fig. 4a), both of which contribute to the shear loss across the vapor layer thickness. We express the loss of shear[34] in the horizontal direction as $\sim \mu_v \bar{u}_r / (\delta + H)^2$, where $\mu_v$ is the viscosity of vapor, $\bar{u}_r$ is the mean outwards velocity of evaporating vapor along the overall flow path expressed as $\sim \left[(T_{sub} - T_{sat}) k_v r_c\right] / \left[\delta h_{fg} \rho_v (\delta + H)\right]$ (Methods). Here $T_{sub} - T_{sat}$ is the difference between the back temperature of sample and the saturation temperature of water, $r_c$ is the radius of contact patch, whereas $k_v$, $h_{fg}$, $\rho_v$ are the thermal conductivity of vapor, the latent heat of vaporization of water, and the density of vapor, respectively. In order to calculate the additional shear loss contributed by post arrays, we used the parallel circuit concept and the resistances of the outward vapor flowing above and inside the post arrays are represented by $R_1 = 1/A_1$ and $R_2 = 1/A_2$ respectively (Supplementary Fig. 4b and c). Here, $A_1 = 2\pi r_c \delta$ and $A_2 = 2\pi r_c (1-\phi) H$ are the relative cross sectional area of outward horizontal flow in the layer 1 and layer 2, respectively, and $\phi = \pi D^2 / (4L^2)$ is the solid fraction. The mass proportion of the vapor inside the post arrays $\psi$ scales as $\psi \sim R_1/(R_1+R_2) = \{1+\delta/[(1-\phi)H]\}^{-1}$. On the basis of mass balance (Methods), the mean velocity of outward flow in layer 2 is written as $\bar{u}_2 \sim \bar{u}_r \psi (H+\delta)/H \sim \bar{u}_r \psi$. Accordingly, the additional shear loss by post arrays is calculated as $\mu_v \bar{u}_r \psi / L^2$. Thus, the total pressure gradient under droplet is:

$$\frac{\Delta P_v}{r_c} \sim \mu_v \bar{u}_r \left[ \frac{\psi}{L^2} + \frac{1}{(H+\delta)^2} \right] \tag{1}$$

By balancing the hammer pressure $\Delta P_h$ with the vapor pressure $\Delta P_v$, the critical superheat needed for the Leidenfrost state is:

$$T_L - T_{sat} \sim \frac{\Delta P_h \delta h_{fg} \rho_v (\delta + H)}{\mu_v k_v r_c^2} / \left[ \frac{\psi}{L^2} + \frac{1}{(H+\delta)^2} \right] \tag{2}$$

where $T_L$ is the Leidenfrost point. Thus, under identical experimental conditions, the surface with larger $L$ is expected to yield a higher Leidenfrost point. Note that in our model we neglect the dynamic pressure since it is much smaller than the hammer pressure. On the basis of experimental measurements, we compared the superheat ($T_L - T_{sat}$) for surfaces with $L$ = 100 μm, 60 μm, 40 μm and 30 μm at $We$ = 19.3, respectively (Fig. 4a). The superheat for the surface of $L$ = 100 μm was 39.4% higher than that of $L$ = 30 μm surface, which was in good agreement with that predicted by our model (41.5%). It is important to emphasize that without considering the hammer pressure, the superheat predicted by the theoretical model for the surface with $L$ = 100 μm is ~73% smaller than that measured experimentally. This emphasizes the importance of considering the hammer pressure in the theoretical analysis.

Based on our model, under a proper temperature range the uniform wetting state imposed to a droplet at the ambient condition is rectified into two distinct regimes as a result of the reliance of its Leidenfrost point on structural topography. The lateral driving force[40] associated with the asymmetric droplet arises from the integrated value of the Laplace pressure gradient over the contour $s$ of the relative contact line at the retraction stage, which is written as $F = \gamma \int_s (\cos\theta_1 - \cos\theta_2) ds$, where $\theta_1$ and $\theta_2$ are the dynamic receding CAs at the leftmost and rightmost meniscus, respectively. The maximum driving force emerges at $T = 265\,^{\circ}\text{C}$. As plotted in Fig. 4b, $\theta_1$ on the gradient surface at this condition was measured to range between ~ 40° and ~ 55° while $\theta_2$ was generally greater than ~150°. As a result, the interaction between the Buller's droplet-like liquid and water film-like film results in a large lateral driving force that propels the entire droplet towards the sticky boiling region. By contrast, at the lower or higher temperature range of the MBL regime, the droplet was primarily dominated by a uniform wetting state (CB or Leidenfrost), and hence the wetting contrast between liquid fronts and the lateral driving force for effective vectoring was reduced. Such asymmetry was also observed for the retraction velocity. As shown in Fig. 4c, the maximum droplet retraction velocity at the liquid meniscus at $T = 265\,^{\circ}\text{C}$ is ~ 0.6 m/s, which is in good agreement with that observed on a superhydrophobic surface[42, 50] ($V_{max} = \sqrt{2\gamma/(\rho h)}$). When the liquid in the Leidenfrost regime was channeled to the CB regime, the contact line velocity was dramatically reduced due to the explosive boiling, as evidenced by our measurement as shown in Fig. 4c. By contrast, the receding CAs and contact line velocities in two moving fronts of the Leidenfrost droplet were almost identical, without contributing to the net directional motion of the droplet (Fig. 4d).

In a broad perspective, the exploration of how structural topography and phase change process mediates the droplet dynamics at high temperature represents an important advance in our understanding of multiphase wetting phenomenon and enables the rational design of various surfaces for multifunctional applications. Figure 5a shows the total time ($t_e$) required for the complete evaporation of an impacting droplet on surfaces as a function of time. In our experiments, the substrate temperature varied between 140 °C and 380 °C and the *We* was kept at 19.3. Four surfaces with uniform post spacing (*L* = 100, 60, 40, and 30 µm) were tested, respectively. At the lower temperature ranges (*T* < 210 °C), the evaporation time for a CB droplet was ~ 0.5 s as a result of explosive boiling. However, the evaporation time in the Leidenfrost condition was two orders of magnitude larger than that in the CB regime. On the basis of measured evaporation time, the average heat transfer coefficient (*h*) was estimated as $h = (\rho h_{fg} V_o) / [A_c (T_{sub} - T_{sat}) t_e]$, where $V_o$ is the initial volume of the water droplet and $A_c$ is the average contact area. Thus, for a constant *T* = 265 °C, the heat transfer coefficient of the boiling droplet on the surface with *L* = 100 µm was calculated over 1460 W/(m² · °C), which was 13.3-fold larger than that of the Leidenfrost droplet on the surface with *L* = 30 µm. Thus, the manifestation of the preferential and rapid motion through the breaking of wetting symmetry in the MBL regime enables an efficient energy exchange between the impacting liquid and hot solid, promoting the energy efficiency, security, and stability of high temperature thermal systems. Otherwise, the establishment of a continuous vapor layer is particularly detrimental in certain contexts, such as in nuclear power plants and high power electronics devices.

**Methods**

**Sample preparation.** The asymmetric surface was fabricated based on 525 µm thick silicon wafer with 1 µm silicon dioxide ($SiO_2$) layer. First, patterned $SiO_2$ layer was first created by using the standard photolithography, followed by oxide dry etching. The silicon oxide pattern serves as a mask for the deep reactive ion etching (DRIE). The DRIE process includes cyclic passivation and etching modes in which $C_4F_8$ and $SF_6$ are used as the reactants, respectively. In this study, the inductively coupled plasma DRIE system (Surface Technology Systems, UK) was used as the etcher. The coil power was set at ~750 W. The chamber pressure and temperature was kept at ~ 94 mTorr and ~ 20 °C. In the passivation cycle, the $C_4F_8$ flow rate was ~ 95 sccm. In the etching cycle, the $SF_6$ flow rate was ~ 140 sccm and platen power was set at ~ 12 W. After the dry etching of micropost arrays, the whole wafer was immersed into the piranha solution (3:1 mixture of $H_2SO_4$ and $H_2O_2$ at 120 °C) for 10 minutes to remove the polymer deposited on the surface during the dry etching stage. The posts are ~ 20 μm in diameter, ~ 20 μm in height and the center-to-center spacing varies from 25 μm to 105 μm.

**Experiment at high temperature.** During the measurement, the gradient sample was taped onto a stainless steel holder, inside which one temperature probe and two cartridge heaters (CIR-2036/240V, Omega) were inserted to control the temperature of the holder. In addition, the temperature of the silicon substrate was measured with the temperature sensor patterned on the back surface of the silicon substrate.

**Leidenfrost point analysis.** The evaporation of water droplet is mainly caused by the large amount of heat conducted from superheated substrate. Thus, the mean velocity of evaporating vapor escaping downward scale as $\bar{u}_e \sim (T_{sub} - T_{sat})/(AR_{th}h_{fg}\rho_v)$, with $A$ being the unit contact area ($A = L^2$) and $R_{th}$ being the overall thermal resistance per unit area. Moreover, the thermal conductivity of silicon is ~ 4000 times larger than vapor, which means that the thermal resistance of vapor layer is still one magnitude larger than the total resistance of silicon base (~505 μm in thickness) and silicon post. Hence, the thermal resistance is primarily governed by the vapor layer above the post arrays, and $R_{th} \approx \delta/(L^2 k_v)$, where $k_v$ is the thermal conductivity of vapor. The downward following vapor finally escapes outward. As for the contact patch with radius $r_c$, the mass balance between the evaporating vapor and outward escaping vapor can be expressed as $2\pi r_c (H+\delta)\bar{u}_r = \pi r_c^2 \bar{u}_e$, thus we got the mean velocity of the overall outward vapor flow $\bar{u}_r \sim [(T_{sub} - T_{sat})r_c k_v]/[\delta h_{fg} \rho_v (H+\delta)]$. In addition, the downward flow rate of vapor in the layer 2 is $\bar{u}_e \psi$. On the basis of mass balance between the downward and outward vapor flow in layer 2, we have $2\pi r_c H \bar{u}_2 = \pi r_c^2 \psi \bar{u}_e$ and $\bar{u}_2 \sim \bar{u}_r \psi (H+\delta)/H$.

**Evaporation experiment.** A thin copper wire with diameter of 80 μm was fixed at the droplet impact point, so that the droplet can be trapped on the sample after collision. Besides, the droplet kept a constant contact with the substrate considering its weight and the small size of the copper wire. The overall evaporation time was recorded by high-speed camera under a frame rate of 60 fps. The diameter of contact base at each moment was measured from the video using Image J. At the lower temperature, the droplet totally wetted the substrate and the contact area was calculated by Wenzel model. By contrast, at the higher temperature, the approximate contact area was calculated based on the Cassie model since a partial or uniform vapor layer was formed between the droplet and substrate.

**Acknowledgements**

This work was supported by the RGC Grant (No. 11213414), the National Natural Science Foundation of China (No. 51475401).

**Author contributions**

**Additional information**

**Supplementary Information** is available in the online version of the paper. Reprints and permissions information is available online at www.nature.com/reprints. Correspondence and requests for materials should be addressed to Z.W. (zuanwang@cityu.edu.hk), and S. Y. (meshyao@ust.hk) or mkc4@lehigh.edu (M.C.)

**Competing financial interests**

The authors declare no competing financial interests.

**Figures captions**

**Figure 1| Asymmetric surface and wetting property characterization.** (**a**) Scanning Electron Microscope (SEM) image of the gradient surface with increasing post-to-post spacing $L$ varying from 25 µm to 105 µm with the step increase of $L$ setting at 5 µm. The patterned microposts have constant diameter of 20 µm and height of 20 µm. (**b**) Snapshot showing the equilibrium state of a droplet on the gradient surface at the ambient condition,

with a slight shift in the center of mass by 0.27 mm.

**Figure 2| Droplet dynamics at the high temperature.** (**a**) Selected snapshots showing the preferential motion of an impinging droplet with $We$ = 19.3 on the asymmetric surface. The left and right columns correspond to $T$ = 245 °C and 265 °C, respectively. The droplet displayed a mixed Boiling-Leidenfrost (MBL) regime. (**b**) Optical image of vapor-liquid-solid interfaces underneath the droplet shows that the liquid penetrated into sparse posts whereas levitated above dense posts. (**c**) Schematic depiction of the spore ejection process in basidiomycetes, which is similar to the droplet vectoring. Briefly, a water droplet (or Buller's droplet, blue color) nucleates at the base of the spore (shaded region). When growing large enough, the droplet merges with the liquid film at the surface of the spore, generating a significant momentum which finally ejects the whole spore. (**d**) The inhibition of preferential motion of the boiling ($T$ = 225 °C) and Leidenfrost droplet ($T$ = 310 °C). In both regimes, the droplet preserved left–right symmetry and no preferential motion was observed.

**Figure 3| Phase diagram.** (**a**) The variation of the maximum displacement factor $k$ of droplets under various Webber numbers as a function of temperature. Filled symbols denote the occurrence of droplet bouncing while open symbols denote no bouncing. (**b**) The variation of the vectoring time for a droplet to efficiently escape from the Leidenfrost regime.

**Figure 4| Dependence of Leidenfrost point on structural roughness.** (**a**) The effect of pillar-to-pillar spacing on the superheat for the Leidenfrost condition. Four control surfaces had constant pillar diameter ($D$ = 20 μm) and height ($H$ = 20 μm), but varying pillar-to-pillar spacing ($L$ = 100 μm, 60 μm, 40 μm and 30 μm). The experimental results were in good

agreement with those predicted by the analytical model. (**b**) The variation of dynamic receding angles of droplets in the Leidenfrost (red, $T = 310\ ^oC$) and MBL (black, $T = 265\ ^oC$) regimes, respectively. The Leidenfrost droplet had almost identical leftmost (solid line) and rightmost (dashed line) receding angle. By contrast, the MBL droplet had a large contrast in two receding angles. (**c, d**) Time evolution of retraction velocities at the leftmost front (red line) and rightmost front (black line) under different temperatures (**c**, $T = 265\ ^oC$, **d**, $T = 310\ ^oC$). The shaded region corresponds to the transition state from the Leidenfrost regime to the CB regime. The contact line velocity in the Leidenfrost regime was much larger than that in the CB regime. By contrast, the contact line velocities in two moving fronts of a Leidenfrost droplet were almost identical.

**Figure 5| Effect of the preferential droplet motion on the heat transfer coefficient.** (**a**) The variation of the evaporation time for an impinging droplet to be fully evaporated under different temperatures (or wetting states). Four control samples with post-to-post spacing of L= 100 μm, 60 μm, 40 μm and 30 μm were tested, respectively. At the lower temperature range ($T < 210\ ^oC$), the evaporation completed at very short timescale (~0.5 s) due to extreme boiling whereas the evaporation time in the Leidenfrost condition was two orders of magnitude larger than that in the CB regime. (**b**) Comparison of heat transfer coefficients for droplets in different wetting states. The heat transfer coefficient of a boiling droplet on the surface with $L = 100$ μm was calculated over 1460 W/(m$^2 \cdot\ ^oC$), which was much larger than that of a Leidenfrost droplet on the surface with $L = 30$μm (109 W/(m$^2 \cdot\ ^oC$)). Thus, the obviation of unwanted Leidenfrost regime dramatically increased the energy efficiency.

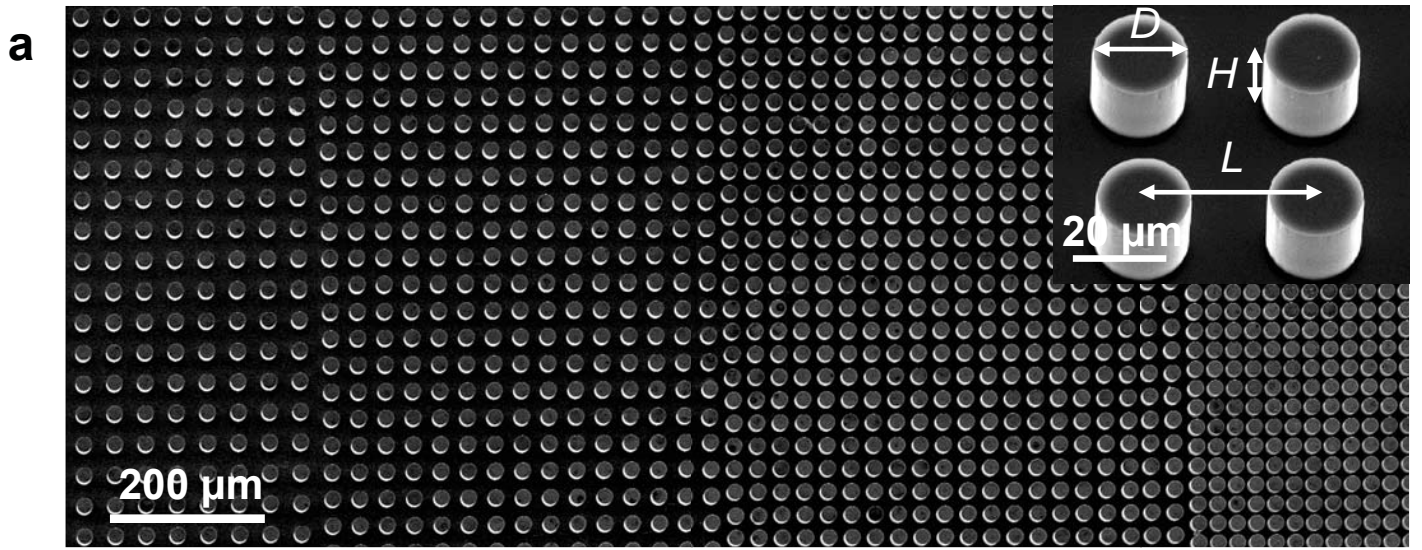

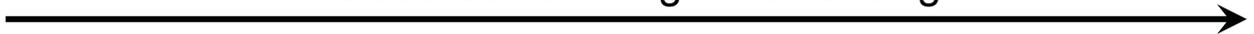

Static Contact Angle Decreasing

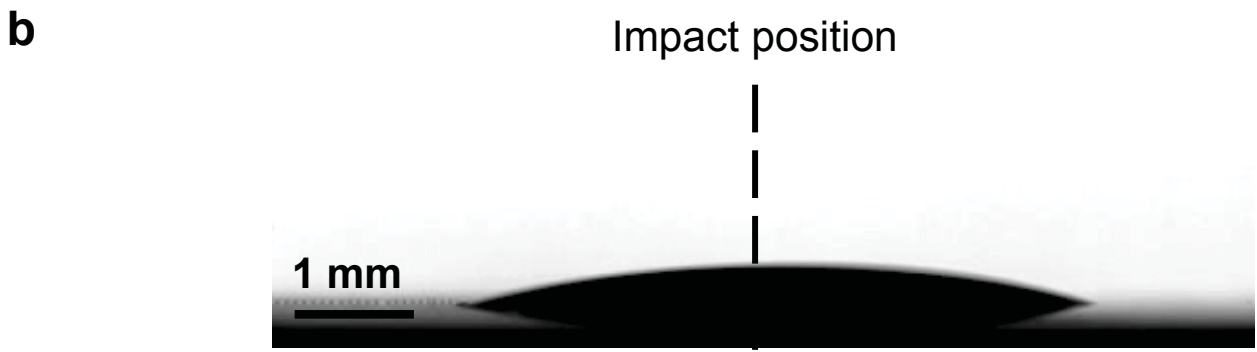

Impact position

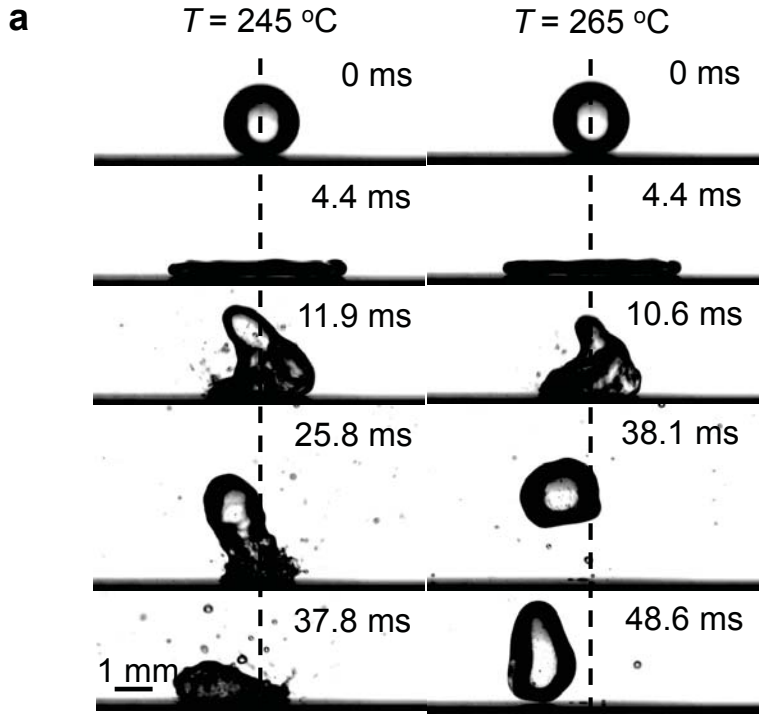

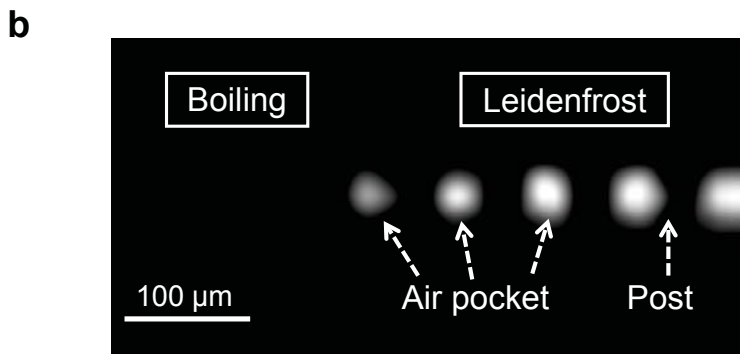

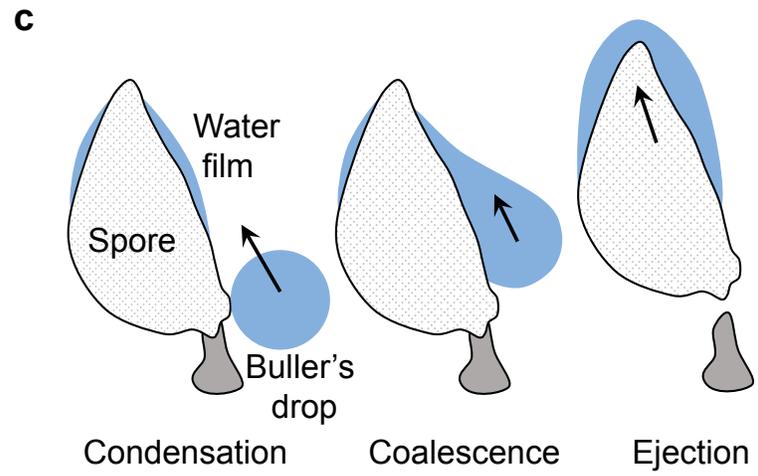

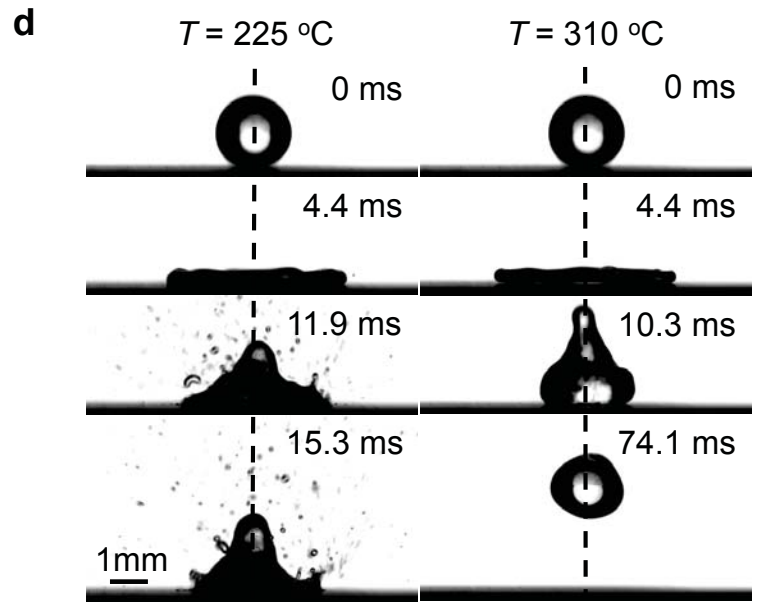

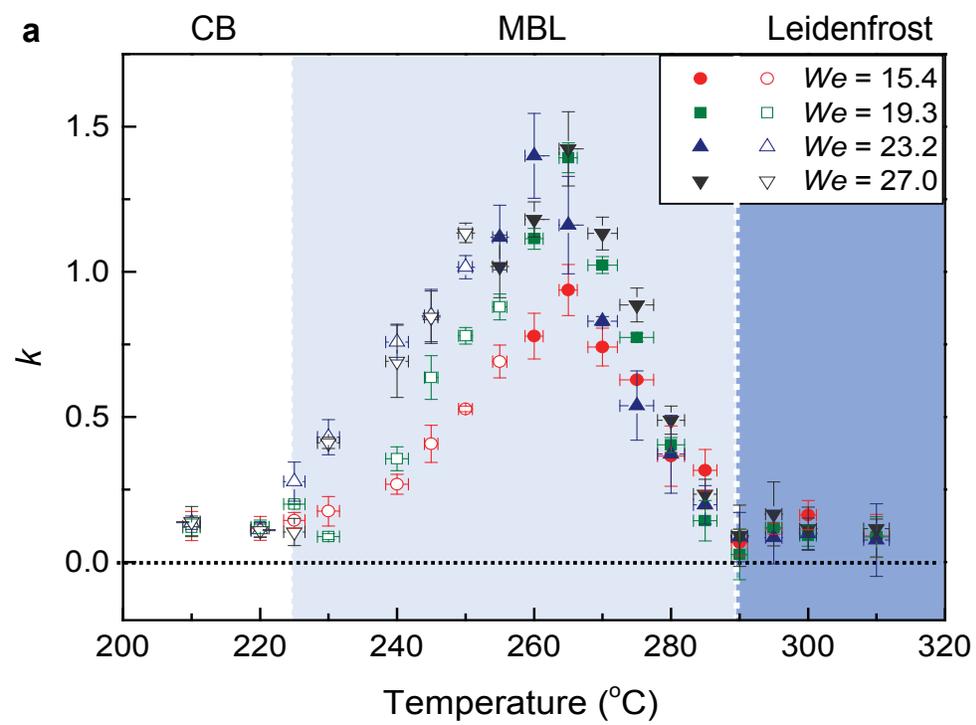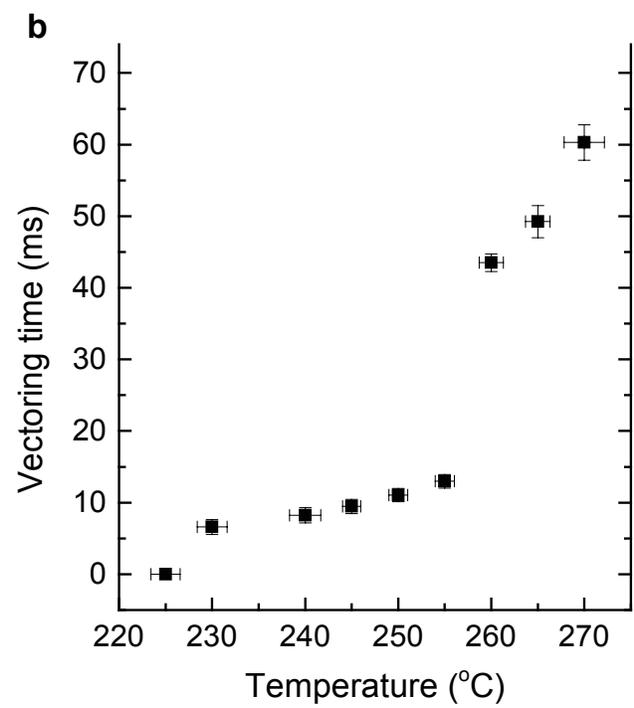

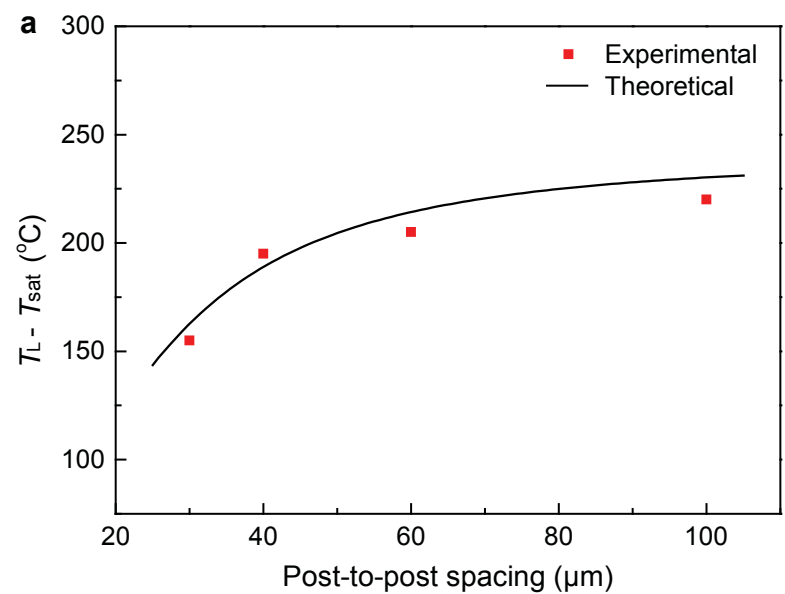
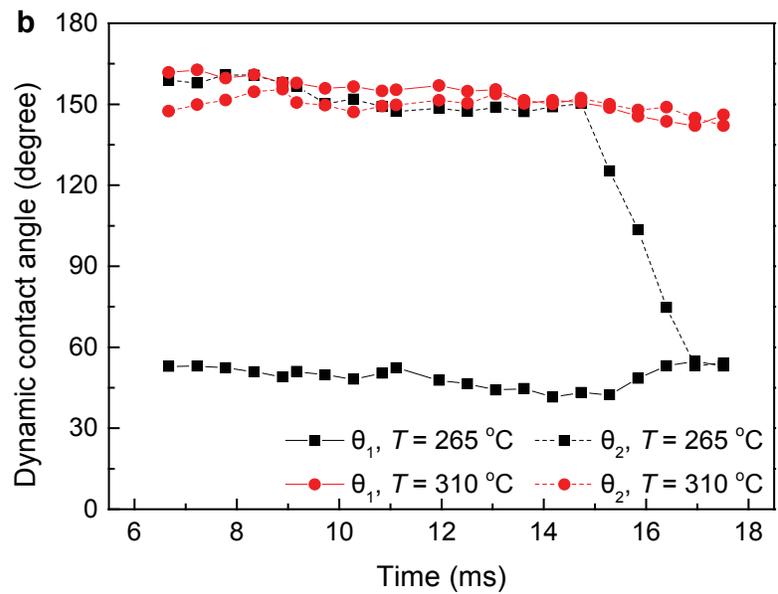
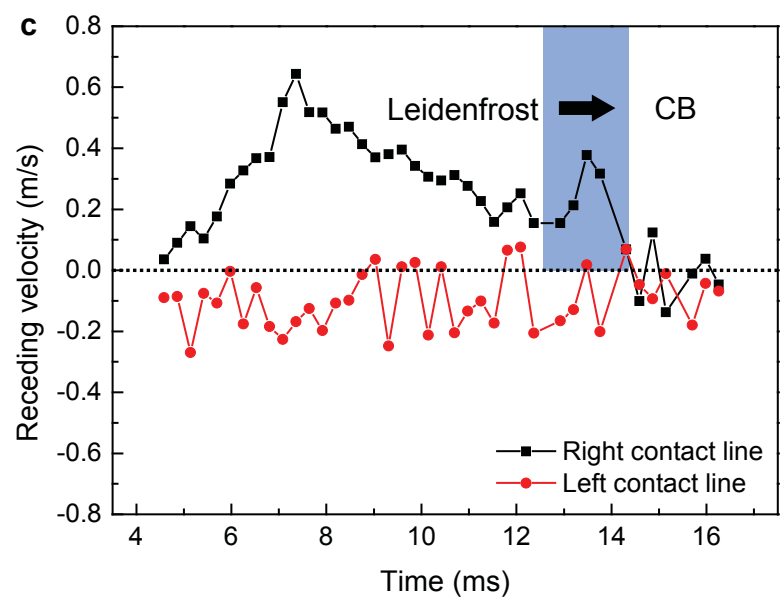
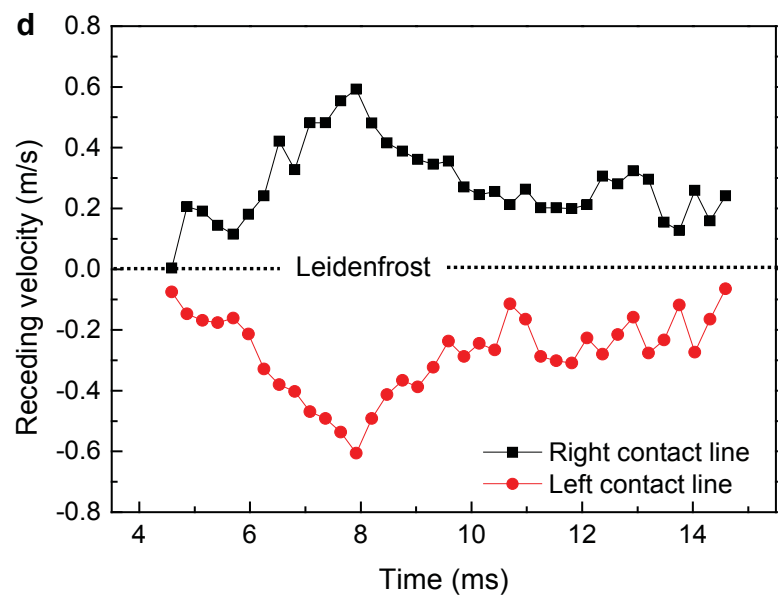

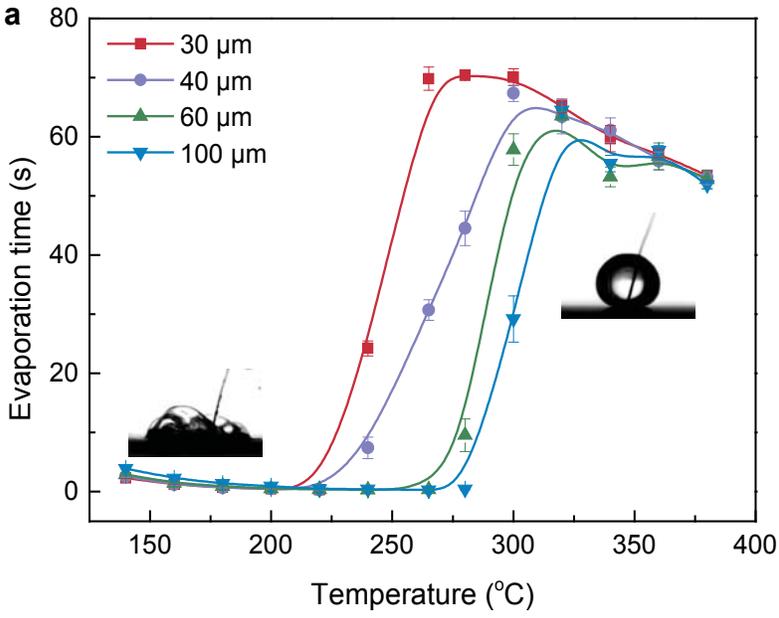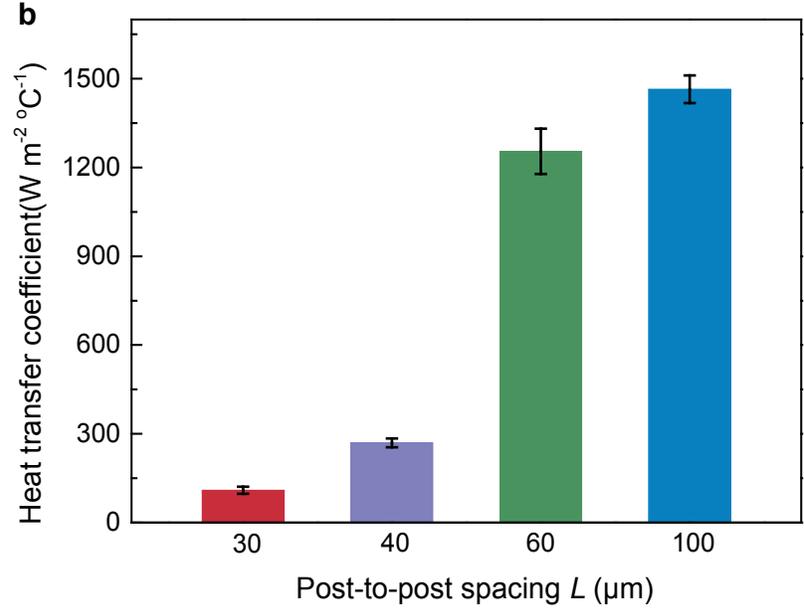